# Bayesian Inference for Structural Vector Autoregressions Identified by Markov-Switching Heteroskedasticity

Helmut Lütkepohl[a], Tomasz Woźniak [☆][b],[*]

[a]*DIW Berlin and Freie Universität Berlin*
[b]*University of Melbourne*

**Abstract**

In this study, Bayesian inference is developed for structural vector autoregressive models in which the structural parameters are identified via Markov-switching heteroskedasticity. In such a model, restrictions that are just-identifying in the homoskedastic case, become over-identifying and can be tested. A set of parametric restrictions is derived under which the structural matrix is globally or partially identified and a Savage-Dickey density ratio is used to assess the validity of the identification conditions. The latter is facilitated by analytical derivations that make the computations fast and numerical standard errors small. As an empirical example, monetary models are compared using heteroskedasticity as an additional device for identification. The empirical results support models with money in the interest rate reaction function.

[☆]The authors acknowledge an excellent research assistance by Yang Song. They thank Dmitri Kulikov, Aleksei Netšunajev, Michele Piffer, Timo Teräsvirta, Luis Uzeda, and Benjamin Wong for the discussions of the manuscript of the paper. They are grateful also to the participants of the seminars at the Monash University, Technische Universität Dortmund, Bank of Canada, Queensland University of Technology, Reserve Bank of New Zealand, Australian National University, and the University of Helsinki, as well as of the presentations during the Workshop on Macroeconomic Research at the Cracow University of Economics, 24th International Conference Computing in Economics and Finance, Rimini Bayesian Econometrics Workshop, Workshop on Structural VAR Models at the Queen Mary University of London, NBER-NSF Seminar on Bayesian Inference in Econometrics and Statistics 2018, Bayesian Analysis and Modeling Summer Workshop 2017, Sydney Time Series & Forecasting Symposium 2017. The computations in this paper were performed using the Spartan HPC-Cloud Hybrid (see Meade, Lafayette, Sauter & Tosello, 2017) at the University of Melbourne. Helmut Lütkepohl acknowledges the support provided by the Deutsche Forschungsgemeinschaft through the SFB 649 "Economic Risk". Tomasz Woźniak acknowledges the support from the Faculty of Business and Economics at the University of Melbourne through the faculty research grant.

[*]*Corresponding author:* Department of Economics, The University of Melbourne, Level 4, FBE Building, 111 Barry Street, Carlton 3053, Victoria, Australia, *Phone:* +61 3 8344 5310, *Fax:* +61 3 8344 6899, *Email address:* tomasz.wozniak@unimelb.edu.au, *URL:* http://bit.ly/tomaszwozniak.

# 1. Introduction

A central problem in structural vector autoregressive (SVAR) analysis is the identification of the structural parameters or, equivalently, the identification of the structural shocks of interest. The identifying assumptions are often controversial. In order to avoid imposing unnecessarily many restrictions, typically only just-identifying restrictions are formulated. In that case the data are not informative about the validity of the restrictions and they cannot be tested with statistical methods. Moreover, even if over-identifying restrictions are imposed, they can only be tested conditionally on a set of just-identifying restrictions. This state of the art has led some researchers to extract additional identifying information from the statistical properties of the data. Notably heteroskedasticity and conditional heteroskedasticity of the reduced form residuals have been used in this context (Rigobon, 2003). Using such additional information may enable the researcher to make the data speak on the validity of restrictions that cannot be tested in a conventional framework.

One model that has been used repeatedly in applied studies lately to capture heteroskedasticity is based on a latent Markov process that drives the changes in volatility. The model was first proposed by Lanne, Lütkepohl & Maciejowska (2010) for SVAR analysis with identification through heteroskedasticity and it was further developed by Herwartz & Lütkepohl (2014). The SVAR model with Markov-switching heteroskedasticity (SVAR-MSH) is in widespread use (see, e.g., Netšunajev (2013), Lütkepohl & Netšunajev (2014, 2017), Lütkepohl & Velinov (2014), Velinov & Chen (2015), Chen & Netšunajev (2017) and (Kilian & Lütkepohl, 2017, Chapter 14)). Some Bayesian methodology has been developed for its analysis by Kulikov & Netšunajev (2013, 2017), Lanne & Luoto (2016) and Woźniak & Droumaguet (2015). Apart from Woźniak & Droumaguet (2015), all Bayesian approaches base inference for these models on draws from the posterior of the reduced form parameters and transform this output into the posterior draws of the structural model identified through heteroskedasticity. Hence their methodology can only be used to generate posterior draws for just-identified structural parameters which limits its applicability when over-identifying restrictions are of interest. Woźniak & Droumaguet (2015) focus on a locally identified SVAR-MSH model and develop methods for drawing from the posterior of the structural parameters. The posterior distribution of the parameters of a locally identified model is multimodal, however, which allows a statistical model comparison, but severely limits the analysis of the structural parameters.

In the present study, a full Bayesian analysis framework is presented based on a SVAR-MSH model where some or all equations are identified. The setup facilitates both of the objectives mentioned above. We emphasize that our setup allows for the possibility that only some of the structural equations and associated structural shocks are identified. In the SVAR literature it is not uncommon that only the responses to a single shock or a small set of shocks are of interest. For example, in a monetary model, the monetary policy shock is often of primary interest. In that case, it makes sense to focus on the structural parameters associated with that shock only. Our approach allows us to handle that situation even if the other shocks are not properly identified.

Our main additional contributions to the SVAR-MSH literature for identification through heteroskedasticity are as follows: (1) Parametric restrictions for global and partial identification of the model are derived and the model is set up in a form such that the data become informative on the conditions required for identification through heteroskedasticity. Moreover, the model setup facilitates the Bayesian estimation of the structural parameters. (2) A procedure for investigating the restrictions for identification of the structural parameters based on a Savage-Dickey density ratio (SDDR) is proposed. For that purpose, a probability distribution is defined that generalizes the beta, *F*, and compound gamma distributions. Thereby a Bayesian statistical procedure is obtained for investigating partial and global identification of the SVAR-MSH model. An SDDR can also be used for assessing the heteroskedasticity of the structural shocks. (3) A fast Markov Chain Monte Carlo (MCMC) sampler is developed for the posterior distribution of the structural parameters and a method for computing the marginal data density (MDD) is provided which facilitates a full Bayesian model selection.

The methods are illustrated by applying them for an empirical analysis of the role of a Divisia money aggregate in a monetary policy reaction function. In a frequentist SVAR analysis, Belongia & Ireland (2015) find support for the hypothesis that Divisia monetary aggregates are important variables in the monetary policy rule. In their conventional SVAR models without accounting for heteroskedasticity they can only test over-identifying restrictions to validate their hypotheses. Using Belongia & Ireland (2015) as a benchmark, the Bayesian methods developed in the current study for the SVAR-MSH model are applied for a broader statistical analysis of the identifying restrictions even for models that are not identified in Belongia & Ireland's framework. We find evidence that a money aggregate is an important factor determining the monetary policy.

The remainder of this study is organized as follows. The next section presents the basic model framework and derives conditions for identification of the structural parameters. Section 3 discusses the prior assumptions used for the structural parameters. The SDDR procedure for investigating the conditions for identification of the structural parameters obtained from the volatility model is presented in Section 4 and the empirical illustration is discussed in Section 5. Conclusions follow in Section 6 and, finally, the proof of a result regarding identification through heteroskedasticity is given in Appendix Appendix A, the computational details of the Gibbs sampler and the estimation of the marginal data densities are presented in Appendix Appendix B, while Appendix Appendix C contains more details on the distribution used in the SDDR procedure. Additional empirical results on the precision of our estimates are presented in Appendix Appendix D.

## 2. Identified Heteroskedastic Structural Vector Autoregressions

### 2.1. The Model

In this section, a structural VAR model is introduced for the *N*-dimensional vector of observable variables $y_t$ in which the structural shocks are conditionally heteroskedastic. The structural-form model is given by

$$A_0 y_t = \mu + A_1 y_{t-1} + \cdots + A_p y_{t-p} + u_t, \tag{1}$$

where the structural matrix $A_0$ is assumed to be nonsingular with unit diagonal, denoted by $\text{diag}(A_0) = \iota_N$, where $\iota_N$ is an *N*-dimensional vector of ones. In other words, there is one equation for each variable. The quantity $\mu$ is an *N*-dimensional vector of constant terms, $A_1, \ldots, A_p$ denote $N \times N$ autoregressive slope coefficient matrices, and $u_t$ is a contemporaneously and serially uncorrelated structural error term. The variances of the structural errors are assumed to change over time according to a latent process $s_t$, $t \in \{1, \ldots, T\}$, and the variance of $u_{it}$ conditional on the state $s_t$ is denoted as $\lambda_{s_t,i}$. Moreover, conditionally on $s_t$, the structural errors are assumed to be normally distributed with mean vector zero and diagonal covariance matrix,

$$u_t | s_t \sim \mathcal{N}(\mathbf{0}, \text{diag}(\lambda_{s_t})), \quad t \in \{1, \ldots, T\}, \tag{2}$$

where $\lambda_{s_t} = (\lambda_{s_t,1}, \ldots, \lambda_{s_t,N})'$ is an $N$-dimensional vector of variances associated with volatility state $s_t$ and $\text{diag}(\lambda_{s_t})$ denotes a diagonal matrix with main diagonal given by $\lambda_{s_t}$. Later in this section, the heteroskedasticity of the data is used to identify the structural matrix $A_0$.

The process $s_t$ for each $t$ can take a discrete number of values, $s_t \in \{1, \ldots, M\}$. In the current study it is assumed to be an unobservable Markov process that defines a Markov-switching model as proposed by Lanne et al. (2010). In principle, the uniqueness restrictions and their Bayesian verification procedure can also be used for other types of processes $s_t$ that describe the volatility changes.

The properties of the Markov process, $s_t$, considered in the current study are fully governed by an $M \times M$ matrix of transition probabilities $\mathbf{P}$. The $(i,j)^{\text{th}}$ element of $\mathbf{P}$ is the probability of switching to state $j$ at time $t$, given that at time $t-1$ the process is in state $i$, $p_{ij} = \Pr[s_t = j | s_{t-1} = i]$, for $i, j \in \{1, \ldots, M\}$, and $\sum_{j=1}^{M} p_{ij} = 1$. Since the hidden Markov process has $M$ states, also $M$ vectors of the state-specific structural error variances, $\lambda_1, \ldots, \lambda_M$, have to be estimated. Such a flexible MS heteroskedastic process offers a range of possibilities of modeling particular patterns of changes in volatility in economic data (see Sims, Waggoner & Zha, 2008; Woźniak & Droumaguet, 2015).

The heteroskedastic SVAR model presented so far allows for the statistical identification of all the $N-1$ free elements in the rows of the structural matrix $A_0$, as will be demonstrated shortly. Therefore, the identified rows of the matrix $A_0$ can be estimated in a heteroskedastic structural form model given by equations (1) and (2). Any further restrictions imposed on the identified rows of $A_0$ over-identify the system, and thus, the data are informative about such restrictions.

## 2.2. Identifying $A_0$ via Heteroskedasticity

To see how statistical identification of the model is obtained via heteroskedasticity, it is useful to study the implied reduced-form model and its relation to the structural form. The reduced form of the model is obtained by multiplying the structural-form model in equation (1) by $A_0^{-1}$ from the left. The reduced-form residuals are $\epsilon_t = A_0^{-1} u_t$ such that, for a uniquely determined matrix $A_0$, the structural errors are obtained from the reduced-form residuals as $A_0 \epsilon_t = u_t$. Suppose that the $M$ covariance matrices of the reduced-form residuals are denoted by $\Sigma_m$, $m \in \{1, \ldots, M\}$. Under the current assumptions, where only the variances of the structural errors are state dependent while the VAR structure is time-invariant, there exists a decomposition

$$\Sigma_m = A_0^{-1} \text{diag}(\lambda_m) A_0^{-1\prime}, \quad m \in \{1, \ldots, M\}. \tag{3}$$

The following theorem presents conditions for identification of the rows of $A_0$. A proof is given in Appendix Appendix A.

**Theorem 1.** Let $\Sigma_m$, $m = 1, \ldots, M$, be a sequence of positive definite $N \times N$ matrices and $\Lambda_m = \text{diag}(\lambda_{m,1}, \ldots, \lambda_{m,N})$ be a sequence of $N \times N$ diagonal matrices with positive diagonal elements. Suppose there exists a nonsingular $N \times N$ matrix $A_0$ with unit main diagonal such that

$$\Sigma_m = A_0^{-1} \Lambda_m A_0^{-1\prime}, \quad m = 1, \ldots, M. \tag{4}$$

Let $\boldsymbol{\omega}_i = (\lambda_{2,i}/\lambda_{1,i}, \ldots, \lambda_{M,i}/\lambda_{1,i})$ be an $(M-1)$-dimensional vector of variances relative to state 1. Then the $k^{\text{th}}$ row of $A_0$ is unique if $\boldsymbol{\omega}_k \neq \boldsymbol{\omega}_i \;\forall i \in \{1, \ldots, N\} \setminus \{k\}$.

The theorem provides a general result on the identification of a single equation through heteroskedasticity. It shows that a structural equation and, hence, the corresponding structural shock is identified if the associated sequence of variances is not proportional to the variance sequences of any of the other shocks. For example, if there are just two volatility states and all variances change proportionally, that is, for some scalar $c$, $\lambda_1 = c\lambda_2$, then $\lambda_{2,i}/\lambda_{1,i} = \lambda_{2,j}/\lambda_{1,j}$, so that the conditions of Theorem 1 are not satisfied.

Theorem 1 implies that $A_0$ is globally identified if the vectors of relative variances $\omega_i$, $i \in \{1, \ldots, N\}$, are all distinct. We summarize this result in the following corollary for future reference.

**Corollary 1.** Under the assumptions of Theorem 1, if $\omega_i \neq \omega_j \ \forall \ i, j \in \{1, \ldots, N\}$, $i \neq j$, then $A_0$ is unique.

Corollary 1 implies that full identification may be obtained even if one of the structural shocks is homoskedastic. For example, in a two-dimensional model the conditions of Corollary 1 are satisfied if the first variance components $\lambda_{1,1}$ and $\lambda_{2,1}$ are equal ($\lambda_{1,1} = \lambda_{2,1}$) as long as $\lambda_{1,2}$ and $\lambda_{2,2}$ are distict.

The global identification condition for $A_0$ in Corollary 1 is an advantage of the present model setup relative to the typical setup used in the related literature on identification via heteroskedasticity. In that literature, a so-called *B*-model is typically used with locally identified shocks, which are unique up to sign and ordering only (see Lütkepohl, 2005). More precisely, the structural errors $u_t$ are assumed to be related to the reduced-form residuals $\epsilon_t$ as

$$u_t = B^{-1}\epsilon_t$$

such that we get a reduced-form covariance decomposition

$$\Sigma_1 = BB', \quad \Sigma_m = B \operatorname{diag}(\lambda_m^*) B', \quad m = 2, \ldots, M. \tag{5}$$

The matrix $B$ has a direct interpretation as the matrix of impact effects of the shocks on the variables. No restrictions are imposed on the main diagonal of $B$ and local uniqueness of $B$ is obtained by normalizing the variances of the structural shocks associated with the first volatility state, that is, $u_t|(s_t = 1) \sim \mathcal{N}(0, I_N)$. The conditions for local uniqueness of this decomposition for any number of states $M$ are derived in Lanne et al. (2010).

While such local identification results are sufficient for asymptotic theory in a frequentist framework, they are not convenient for Bayesian analysis because they complicate simulating posterior distributions. Thus, the setup of Corollary 1, with restricted diagonal elements of $A_0$ and a specific variance sequence associated with each equation, is particularly useful for Bayesian analysis. Moreover, estimation and inference of the unrestricted parameters of the matrix $A_0$ in the current model is separated from the scaling problem associated with the label switching of heteroskedastic states. In effect, the likelihood function and the posterior distribution have more regular shapes with fewer modes (see Woźniak & Droumaguet, 2015, for the detailed analysis of the impact of label switching of the heteroskedastic states on the shape of the posterior distribution).

Conditions for full identification could be formulated equivalently for parametrisation (5). In fact, instead of normalizing the diagonal elements of $A_0$, one could normalize the diagonal elements of $B$ to obtain global identification. Such a normalization amounts to assuming that the $k^{\text{th}}$ shock has unit instantaneous impact on the $k^{\text{th}}$ variable. That condition is used by Stock &

Watson (2016) who list several of its advantages. However, a potential drawback is that such a normalization requires knowledge that the $k^{\text{th}}$ shock has a nonzero impact effect on the $k^{\text{th}}$ variable which may not be obvious in some situations. Thus, we prefer to work with a normalized $A_0$ matrix.

The advantage of the conditions given in Theorem 1 and Corollary 1 for the variances in parametrization (3) is that they can be investigated by statistical methods because the data are informative about them. If the conditions for full identification in Corollary 1 are not satisfied, the changing volatility may still offer some additional identifying information that implies sufficient curvature in the likelihood and, hence, in the posterior, to enable the data to discriminate between competing economic models.

Note that the identification of the matrix $A_0$ using heteroskedasticity is only a statistical identification that allows to estimate all or some of the elements of this matrix without imposing any further restrictions on the model. For any identified shock, the structural impulse response functions can be computed. However, the structural-form errors do not have economic interpretations as such. In order to call any of the structural shocks, say a monetary policy shock, economic reasoning needs to be imposed. Still, it is useful to exploit such an identification of the shocks as it opens up the possibility for testing any further restrictions imposed on the model on the basis of economic considerations.

### 2.3. Imposing Restrictions on the Matrix $A_0$

In order to obtain a flexible framework that facilitates the estimation of models with unrestricted or restricted matrix $A_0$, the approach proposed by Amisano & Giannini (1997), used also by Canova & Pérez Forero (2015) is helpful. Let the $r \times 1$ vector $\alpha$ collect all of the unrestricted elements of the matrix $A_0$ column by column. Then we impose restrictions on the structural matrix $A_0$ by setting

$$\text{vec}(A_0) = Q\alpha + q, \tag{6}$$

where $Q$ and $q$ are respectively an $N^2 \times r$ matrix and an $N^2 \times 1$ vector. Typically the elements of $Q$ and $q$ will be zeros and ones if zero restrictions are imposed on the off-diagonal elements of $A_0$ in addition to the restrictions due to normalizing the main diagonal.

## 3. Prior Distributions for Bayesian Analysis

To facilitate the inference on the restrictions for the uniqueness of the rows of the matrix $A_0$ with ones on the main diagonal we estimate state-specific variances of the structural shocks in a parametrization that includes the variances of the structural shocks in the first state, $\lambda_1$, and $M-1$ vectors of relative variances, $\omega_m = [\lambda_{m,i}/\lambda_{1,i}]$, for states $m \in \{2,\dots,M\}$. Specifying independent inverse gamma 2 distributions ($\mathcal{IG}2$) as prior distributions for the $\omega_m$s, given our assumptions about the distribution of the error terms, leads to the full conditional posterior distributions for these parameters being of the same type.[1] This setup is the basis for feasible computations of the SDDRs for the uniqueness conditions that are specified for relative variances. The choice of the parametrization, marginal prior and full conditional posterior distributions for the $\omega_m$s makes our framework general. Note that it does not dependent on the inference on the latent state variable $s_t$. The details of Bayesian assessment of the uniqueness restrictions are given in Section 4.

---

[1] For the definition of the distribution, its properties, and the random numbers sampling algorithm see Bauwens, Richard & Lubrano (1999, Appendix B).

The variances of the structural shocks in the first state, $\lambda_{1,n}$, are *a priori* independently distributed as $\mathcal{IG}2$ with parameters $\underline{a}_\lambda$ and $\underline{b}_\lambda$ set to 1 which makes the distribution quite spread out over a wide range. In fact, under this assumption the first and second moments of $\lambda_{1,n}$ may be infinite which limits the impact of the prior distribution on the posterior distribution.

The prior of each of the relative variances of structural shocks $\omega_{m,n}$, for $m \in \{2,\dots,M\}$, follows independently an $\mathcal{IG}2$ distribution with parameters $\underline{a}_\omega = 1$ and $\underline{b}_\omega = \underline{a}_\omega + 2$, which ensures that the mode of the prior distribution is located at 1. This assumption implies that the state-specific variances for states $2,\dots,M$ have prior distributions similar to those in Woźniak & Droumaguet (2015). At the mode of the prior distribution there is no heteroskedasticity and hence the rows of $A_0$ are not uniquely identified.

The prior distribution for the unrestricted elements of the matrix $A_0$ collected in the vector $\alpha$, conditionally on hyper-parameter $\gamma_\alpha$, is a normal distribution with mean vector zero and a diagonal covariance matrix $\gamma_\alpha I_r$. To avoid making the prior more restrictive for some elements of $A_0$ than for others one could make sure that the variables entering the model have similar orders of magnitude. In macro models this is typically not a problem because many variables enter in logs or rates of change. The hyper-parameter $\gamma_\alpha$ is interpreted as the level of shrinkage imposed on the structural parameters $\alpha$ and is also estimated. For that purpose, we define the marginal prior distribution of $\gamma_\alpha$ to be $\mathcal{IG}2$ with parameters $\underline{a}$ and $\underline{b}$ set to 1.

The conditional prior distribution of the variable-specific constant term, $\mu_n$, $n \in \{1,\dots,N\}$, given a constant term specific hyper-parameter $\gamma_\mu$, is a univariate normal distribution with mean zero and variance $\gamma_\mu$. The marginal prior distribution for $\gamma_\mu$ is $\mathcal{IG}2$ with parameters $\underline{a}$ and $\underline{b}$ set to 1.

To specify the prior distribution of the structural VAR slope parameters $\beta = [A_1,\dots,A_p]$, let $\underline{P} = \begin{bmatrix} D & \mathbf{0}_{N\times N(p-1)} \end{bmatrix}$, where $D$ is an $N \times N$ diagonal matrix. Typically the diagonal elements of the matrix $D$ are zeros for stationary variables and ones for persistent variables, as in the Minnesota prior, but they could also be other known quantities. Then the conditional prior distribution of the equation-specific autoregressive parameters, $\beta_n = [A_{1.n},\dots,A_{p.n}]$, where $A_{l.n}$ is the $n^{\text{th}}$ row of matrix $A_l$ for $l \in \{1,\dots,p\}$, is a $pN$-variate normal distribution. It is conditioned on an autoregressive hyper-parameter $\gamma_\beta$ and the $n^{\text{th}}$ row of $A_0$, denoted by $A_{0.n}$. Its prior mean is equal to $A_{0.n}\underline{P}$ and its prior covariance matrix is equal to $\gamma_\beta \underline{H}$. The diagonal matrix $\underline{H}$ has the main diagonal set to the vector $\left((1^2)^{-1}\iota'_N, (2^2)^{-1}\iota'_N, \dots, (p^2)^{-1}\iota'_N\right)'$, and thus it allows to impose a decaying pattern of prior variances for the subsequent lags as in the Minnesota prior of Doan, Litterman & Sims (1983). The prior distribution for $\gamma_\beta$ is $\mathcal{IG}2$ with parameters $\underline{a}$ and $\underline{b}$ set to 1.

Finally, denote by $\mathbf{P}_m$ the $m^{\text{th}}$ row of the transition matrix $\mathbf{P}$. The prior distributions for the rows of the transition probabilities matrix, $\mathbf{P}_m$, are set independently for each row and are given by $M$-dimensional Dirichlet distributions ($\mathcal{D}_M$) as in Woźniak & Droumaguet (2015). The parameters of these distributions, $e_{m,k}$, for $k \in \{1,\dots,M\}$, are all set to 1 except the parameters corresponding to the diagonal elements of the matrix $\mathbf{P}$ of transition probabilities, denoted by $e_{m,m}$, which are set to 10. This choice expresses the prior assumption that the volatility states are persistent over time.

To summarize, the prior specification takes the following form:

$$p(\theta) = p(\gamma_\alpha)p(\gamma_\mu)p(\gamma_\beta)p(\alpha|\gamma_\alpha)\left(\prod_{m=1}^{M}p(\mathbf{P}_m)\right)$$
$$\times \left(\prod_{n=1}^{N}p(\mu_n|\gamma_\mu)p(\beta_n|A_{0.n},\gamma_\beta)p(\lambda_{1.n})\left(\prod_{m=2}^{M}p(\omega_{m.n})\right)\right), \quad (7)$$

where the specific prior distributions are:

$$\mu_n|\gamma_\mu \sim \mathcal{N}\left(0,\gamma_\mu\right)$$
$$\beta'_n|A_{0.n},\gamma_\beta \sim \mathcal{N}_{pN}\left(A_{0.n}\underline{P},\gamma_\beta\underline{H}\right)$$
$$\alpha|\gamma_\alpha \sim \mathcal{N}_r\left(\mathbf{0}_r,\gamma_\alpha I_r\right)$$
$$\lambda_{1,n} \sim \mathcal{IG}2\left(\underline{a}_\lambda,\underline{b}_\lambda\right)$$
$$\omega_{\tilde{m},n} \sim \mathcal{IG}2\left(\underline{a}_\omega,\underline{b}_\omega\right)$$
$$\gamma_\alpha \sim \mathcal{IG}2\left(\underline{a},\underline{b}\right)$$
$$\gamma_\mu \sim \mathcal{IG}2\left(\underline{a},\underline{b}\right)$$
$$\gamma_\beta \sim \mathcal{IG}2\left(\underline{a},\underline{b}\right)$$
$$\mathbf{P}_m \sim \mathcal{D}_M\left(\underline{e}_{m1},\ldots,\underline{e}_{mM}\right)$$

for $n \in \{1,\ldots,N\}$, $m \in \{1,\ldots,M\}$ and $\tilde{m} \in \{2,\ldots,M\}$.

The above choice of the prior distributions is practical. Priority is given to distributions that result in convenient and proper full conditional posterior distributions, and therefore, allow for the derivation of an efficient Gibbs sampler that is described in Appendix Appendix B. The hierarchical prior distributions for the constant terms, autoregressive slope parameters, and the structural matrix constitute a flexible framework in which the impact of the choice of the hyper-parameters of the prior distribution on inference is reduced, in line with Giannone, Lenza & Primiceri (2015).

## 4. Bayesian Assessment of Identification Conditions and Heteroskedasticity

In this section, we propose to use the Savage-Dickey Density Ratio (SDDR) (see Verdinelli & Wasserman, 1995, and references therein) to verify the identification conditions for the structural model considered in this paper. The SDDR is one of the methods to compute a Bayes factor. The Bayes factor itself, under the assumption of equal prior probabilities of the competing models, is interpreted as a posterior odds ratio of the model with restrictions versus the unrestricted model. Thus, a large value of the SDDR is evidence in favor of the restriction considered and a small SDDR provides evidence against the restriction.

The main advantage of verifying hypotheses using SDDRs is the small computational cost required relative to alternative inference methods. SDDRs are computed using only the output of the unrestricted model estimation through MCMC techniques. If a probability density function of the restricted (function of) parameters is available, then the computations simplify even further by the application of the Rao-Blackwell tool (see Gelfand & Smith, 1990). The latter provides a marginal density ordinate estimate with a high numerical precision. The SDDR for verification of identification of the SVAR proposed below exhibits all of the features mentioned above. Alternative ways of computing the Bayes factor either are associated with the loss of numerical efficiency or involve estimation of two models, the restricted and the unrestricted ones, as well as estimation of MDDs. Either of the tasks may require significant computational costs.

The identification of SVARs requires sufficient variability in the conditional variances of the structural shocks. The uniqueness of the structural matrix $A_0$ can be assessed by verifying the equality restrictions for specific relative variances, such as:

$$\omega_{m,i} = \omega_{m,j}. \tag{8}$$

If the restriction above holds for some $i$ and $j$ for all $m \in \{2, \ldots, M\}$, then the $i^{\text{th}}$ structural shock cannot be distinguished from the $j^{\text{th}}$ structural shock and the corresponding rows of matrix $A_0$ are not uniquely identified.

Of course, identification through heteroskedasticity requires that there are at least two distinct volatility states. In other words, for heteroskedasticity of structural shock $i$, there must be two distinct variances $\lambda_{m,i}$, $m \in \{1, \ldots, M\}$, which translates to the requirement that at least one of the $\omega_{m,i}$, $m \in \{2, \ldots, M\}$ is not equal to one. Thus, the homoskedasticity of the $i^{\text{th}}$ structural shock is assessed by verifying restrictions

$$\omega_{2,i} = \cdots = \omega_{M,i} = 1. \tag{9}$$

Both of the restrictions (8) and (9) can be verified as the data are informative about these features.

*4.1. Identification Conditions*

We rewrite the restriction in equation (8) as

$$\frac{\omega_{m,i}}{\omega_{m,j}} = 1 \tag{10}$$

and use the SDDR to evaluate its validity. The SDDR for the restriction given in equation (10) is a ratio of the marginal posterior distribution to the marginal prior distribution of the left-hand side of the restriction both evaluated at the restricting value. Formally,

$$SDDR = \frac{p\left(\frac{\omega_{m,i}}{\omega_{m,j}} = 1 | Y\right)}{p\left(\frac{\omega_{m,i}}{\omega_{m,j}} = 1\right)}, \tag{11}$$

where $Y = (y_1, \ldots, y_T)$ denotes the data. Small values of the SDDR provide evidence against the ratio $\frac{\omega_{m,i}}{\omega_{m,j}}$ being 1. Of course, this raises the question how small the SDDR has to be to indicate clear evidence against the restriction. Kass & Raftery (1995) discuss a scale for evaluating the size of the SDDR. We will use that scale in our empirical illustration in Section 5.

The SDDR is particularly suitable for the verification of the identification conditions because it does not require the estimation of the restricted models. Moreover, the SDDR can be easily computed as long as the densities of the full conditional posterior and the prior distributions of $\frac{\omega_{m,i}}{\omega_{m,j}}$ are of known analytical form. We propose a distribution that is useful for such computations in the context of $\mathcal{IG}2$ distributed relative variances of the structural shocks.

**Definition 1 (Inverse Gamma 2 Ratio distribution)** Let $x$ and $y$ be two strictly positive independent random variables distributed according to the following $\mathcal{IG}2$ distributions: $x \sim \mathcal{IG}2(a_1, b_1)$ and $y \sim \mathcal{IG}2(a_2, b_2)$, where $a_1$, $a_2$, $b_1$, and $b_2$ are positive real numbers and the probability density function of the inverse gamma 2 distribution is given by:

$$f_{\mathcal{IG}2}(x; a, b) = \Gamma\left(\frac{a}{2}\right)^{-1} \left(\frac{b}{2}\right)^{\frac{a}{2}} x^{-\frac{a+2}{2}} \exp\left\{-\frac{1}{2}\frac{b}{x}\right\}, \tag{12}$$

where $\Gamma(\cdot)$ denotes the gamma function. Then, the random variable $z$, defined as $z = x/y$, follows the *Inverse Gamma 2 Ratio* ($\mathcal{IG}2\mathcal{R}$) distribution with probability density function given by:

$$f_{\mathcal{IG}2\mathcal{R}}(z; a_1, a_2, b_1, b_2) = B\left(\frac{a_1}{2}, \frac{a_2}{2}\right)^{-1} b_1^{\frac{a_1}{2}} b_2^{\frac{a_2}{2}} z^{\frac{a_2-2}{2}} (b_1 + b_2 z)^{-\frac{a_1+a_2}{2}}, \tag{13}$$

where $B(\cdot, \cdot)$ denotes the beta function. □

It is easy to show that the moments of the Inverse Gamma 2 Ratio distribution are as follows.

**Moments of the $\mathcal{IG}2\mathcal{R}$ distribution.** The expected value and the variance of the $\mathcal{IG}2\mathcal{R}$–distributed random variable $z$ are respectively given by

$$\mathbb{E}[z] = \frac{b_1}{b_2} \frac{a_2}{a_1 - 2} \qquad \text{for} \quad a_1 > 2, \tag{14}$$

$$Var[z] = \frac{2\left(\frac{b_1}{b_2}\right)^2 a_2(a_1 + a_2 - 2)}{(a_1 - 2)^2(a_1 - 4)} \qquad \text{for} \quad a_1 > 4. \tag{15}$$

In general, the $k^{\text{th}}$ order non-central moment of $z$ is given by

$$\mathbb{E}\left[z^k\right] = \left(\frac{b_1}{b_2}\right)^k \frac{B\left(\frac{a_1-2k}{2}, \frac{a_2+2k}{2}\right)}{B\left(\frac{a_1}{2}, \frac{a_2}{2}\right)} \qquad \text{for} \quad a_1 > 2k. \tag{16}$$

□

The density given above generalizes the $F$ distribution that is nested within our distribution family by setting $a_1 = b_1$ and $a_2 = b_2$, as well as the compound gamma distribution derived by Dubey (1970) that is parametrized by three parameters $a_1/2$, $a_2/2$, and $b_1/b_2$.[2] For completeness of the derivations, Appendix C defines the Inverse Gamma 1 Ratio distribution of a random variable that is defined as a ratio of two independent inverse gamma 1-distributed random variables. The probability density function as well as the moments of that distribution are also established. These results may facilitate the computations if one prefers to parametrize the model in terms of the conditional standard deviations instead of conditional variances $\lambda$ and $\omega$.

---

[2] Further generalizations of the $F$, Beta, and compound gamma distributions were proposed by McDonald (1984) and McDonald & Xu (1995). The latter work is particularly relevant for our developments as it proposes the generalizations of the compound gamma distributions parametrized by four and five parameters. Their distributions explicitly nest the compound gamma distribution, however, none of them nests our Inverse Gamma 2 Ratio or the Inverse Gamma 1 Ratio distribution.

In Section 3 it was assumed that the parameters $\omega_{m,i}$ and $\omega_{m,j}$ are *a priori* independently distributed as $\mathcal{IG}2$. In effect, the denominator of the SDDR from equation (11) can be computed by simply evaluating the newly proposed distribution with parameters $a_1 = a_2 = \underline{a}_\omega = 1$ and $b_1 = b_2 = \underline{b}_\omega = 3$ at value $z = 1$.

In Appendix Appendix B it is shown that, given the data, realizations of the Markov process, and other parameters, the relative variances are independently $\mathcal{IG}2$ distributed. We use this feature to compute the numerator of the SDDR, apply the Rao-Blackwell tool of Gelfand & Smith (1990), and obtain

$$\hat{p}\left(\frac{\omega_{m,i}}{\omega_{m,j}} = 1 \Big| Y\right) = \frac{1}{S} \sum_{s=1}^{S} f_{\mathcal{IG}2\mathcal{R}}\left(1; a_{i,m}^{(s)}, a_{j,m}^{(s)}, b_{i,m}^{(s)}, b_{j,m}^{(s)}\right), \tag{17}$$

where $\left\{a_{i,m}^{(s)}, a_{j,m}^{(s)}, b_{i,m}^{(s)}, b_{j,m}^{(s)}\right\}_{s=1}^{S}$ is a sample of $S$ draws from the posterior distribution defined for $n \in \{i, j\}$ as follows:

$$a_{n,m}^{(s)} = \underline{a}_\omega + T_m^{(s)}, \tag{18a}$$

$$b_{n,m}^{(s)} = \underline{b}_\omega + \left(\lambda_{1,n}^{(s)}\right)^{-1} \sum_{t=1}^{T} \left(A_{0,n}^{(s)} y_t - \mu_n^{(s)} - A_{1,n}^{(s)} y_{t-1} - \cdots - A_{p,n}^{(s)} y_{t-p}\right)^2, \tag{18b}$$

where $T_m^{(s)}$ is the number of observations classified as belonging to the $m^{\text{th}}$ state in the $s^{\text{th}}$ iteration of the sampling algorithm.

According to the conditions stated in Section 2, the $j^{\text{th}}$ structural shock may not be identified if all the ratios $\frac{\omega_{m,i}}{\omega_{m,j}}$ are equal to 1. Hence, to establish possible identification problems, we have to investigate whether $\frac{\omega_{m,i}}{\omega_{m,j}} = 1$ holds for all $m \in \{2, \ldots, M\}$. The SDDR can be extended for that purpose. Let $\mathcal{U}_{i,j}$ denote the event that $\frac{\omega_{m,i}}{\omega_{m,j}} = 1$ holds for $m \in \{2, \ldots, M\}$. In such a case, the denominator of the SDDR for $\mathcal{U}_{i,j}$ is computed simply as:

$$\hat{p}\left(\mathcal{U}_{i,j}\right) = \prod_{m=2}^{M} p\left(\frac{\omega_{m,i}}{\omega_{m,j}} = 1\right) = f_{\mathcal{IG}2\mathcal{R}}\left(1; \underline{a}_\omega, \underline{a}_\omega, \underline{b}_\omega, \underline{b}_\omega\right)^{M-1}, \tag{19}$$

where the last equality comes from the assumption of the invariance of the prior distribution with respect to $m$. The SDDR's numerator is computed as:

$$\hat{p}\left(\mathcal{U}_{i,j}\Big|Y\right) = \frac{1}{S} \sum_{s=1}^{S} \prod_{m=2}^{M} f_{\mathcal{IG}2\mathcal{R}}\left(1; a_{i,m}^{(s)}, a_{j,m}^{(s)}, b_{i,m}^{(s)}, b_{j,m}^{(s)}\right). \tag{20}$$

The computations of the SDDRs presented above, given the output from the MCMC estimation, are fast and accurate, which emphasizes the advantages of the current setup. The verification of the identification conditions with the SDDRs requires the prior and full conditional posterior distributions for the relative variances of the structural shocks being $\mathcal{IG}2$- or $\mathcal{IG}1$-distributed.

The setup can be further generalized by assuming hierarchical prior distributions for the relative variances, $\omega_{m,n}$, and is independent on how the state variable $s_t$ is estimated. It is, therefore, easily applicable also to other regime-dependent heteroskedastic processes such as those considered by Woźniak & Droumaguet (2015) and Markov-switching models with time-varying transition probabilities as considered by Sims et al. (2008) and Chen & Netšunajev (2017).

Having a procedure for verifying the identification conditions emphasizes the benefits of applying Bayesian inference in this paper. Note that there does not exist a general, valid frequentist test of such conditions. In Bayesian inference the estimation of a model that is not identified does not pose any theoretical or practical obstacles. Therefore, using a standard way of verifying hypotheses, such as through the SDDR, is straightforward. Still, verification of the uniqueness is essential to understand the SVAR model identified through heteroskedasticity.

*4.2. Homoskedasticity*

If the identification conditions are confirmed, then heteroskedasticity is also established as a by-product. However, one may also be interested in testing the shocks individually or jointly for heteroskedasticity. In a similar way as the SDDR can be used to verify the identification conditions, it can also be used to investigate the heteroskedasticity of the structural shocks. Denote by $\mathcal{H}_i$ the event that the restrictions $\omega_{2,i} = \cdots = \omega_{M,i} = 1$ hold, which is the condition for the homoskedasticity of the $i^{\text{th}}$ shock. The SDDR for assessing this hypothesis is given by

$$SDDR = \frac{p(\mathcal{H}_i|Y)}{p(\mathcal{H}_i)}. \tag{21}$$

The elements of the SDDR in the equation above can be computed easily by

$$\hat{p}(\mathcal{H}_i) = \prod_{m=2}^{M} p(\omega_{m,i} = 1) = f_{IG2}\left(1; \underline{a}_\omega, \underline{b}_\omega\right)^{M-1} \tag{22}$$

and

$$\hat{p}(\mathcal{H}_i|Y) = \prod_{m=2}^{M} p(\omega_{m,i} = 1|Y) = \frac{1}{S}\sum_{s=1}^{S}\prod_{m=2}^{M} f_{IG2}\left(1; a_{i,m}^{(s)}, b_{i,m}^{(s)}\right), \tag{23}$$

where $a_{i,m}^{(s)}$ and $b_{i,m}^{(s)}$ are given in equation (18).

The condition for joint homoskedasticity of several structural shocks can be assessed as well. Let $\mathcal{J}$ be a set of $K \leq N$ indicators that define the considered conjunction of homoskedasticity conditions:

$$\mathcal{J} = \left\{j_i \in \{1, \ldots, N\} \text{ for } i \in \{1, \ldots, K\} : \mathcal{H}_{j_1} \cap \cdots \cap \mathcal{H}_{j_K}\right\}.$$

Then, the joint homoskedasticity condition is denoted by $\mathcal{H} = \mathcal{H}_{j_1} \cap \cdots \cap \mathcal{H}_{j_K}$ and the elements of the SDDR are computed as follows:

$$\hat{p}(\mathcal{H}) = \prod_{i \in \mathcal{J}}\prod_{m=2}^{M} p(\omega_{m,i} = 1) = f_{IG2}\left(1; \underline{a}_\omega, \underline{b}_\omega\right)^{K(M-1)} \tag{24}$$

and

$$\hat{p}\left(\mathcal{H}|Y\right) = \prod_{i\in\mathcal{J}}\prod_{m=2}^{M} p\left(\omega_{m,i} = 1|Y\right) = \frac{1}{S}\sum_{s=1}^{S}\prod_{i\in\mathcal{J}}\prod_{m=2}^{M} f_{\mathcal{IG}2}\left(1; a_{i,m}^{(s)}, b_{i,m}^{(s)}\right). \tag{25}$$

All the computations in this section are facilitated by the fact that $\omega_{m,n}$ are *a priori* as well as conditionally *a posteriori* independent for $m \in \{2, \ldots, M\}$ and $n \in \{1, \ldots, N\}$. The proposed Bayesian SDDR for assessing homoskedasticity can be computed easily given the sample of draws from the posterior distribution and it does not pose any significant theoretical challenges (see, e.g., Frühwirth-Schnatter, 2006).

## 5. Empirical Illustration

### 5.1. Background

In this section we illustrate our Bayesian procedures by applying them to SVAR models that were considered by Belongia & Ireland (2015) to study the role of Divisia monetary aggregates in monetary policy models. These authors find statistical support for the importance of Divisia monetary aggregates in the monetary policy rule. They document these relationships using Divisia measurements of several alternative monetary aggregates.

In this paper, we focus on the particular role of the money aggregate M2 that, when properly represented by a Divisia measure, has the capability of explaining aggregate fluctuations to a large extent, as argued by Barnett (2012). For that purpose, we review a number of identification schemes for the SVAR model some of which have been considered by Belongia & Ireland (2015). We build VAR models for the following six quarterly U.S. variables: $p_t$ - log of GDP deflator, $gdp_t$ - log of real GDP, $cp_t$ - a measure of commodity prices defined as the spot index compiled now by the Commodity Research Bureau and earlier by the Bureau of Labor Statistics, $FF_t$ - federal funds rate, $M_t$ - M2 Divisia monetary aggregate to measure the flow of monetary services and $m_t$ its logarithm, $uc_t$ - user-cost measure, provided by Barnett et al. (2013), that is the price dual to the Divisia monetary aggregate $M_t$. These variables in exactly this order are collected in the vector $y_t$, i.e., $y_t' = (p_t, gdp_t, cp_t, FF_t, m_t, uc_t)$. The series are plotted in Figure 1 for the sample period from 1967Q1 - 2013Q4.[3]

### 5.2. Alternative Identification Schemes

Assuming that there is sufficient heterogeneity in the covariance structure of the VAR model, a full set of shocks can be identified by heteroskedasticity. No further restrictions are needed for $A_0$ in this case. In the following we refer to the model as *unrestricted* if it is identified purely by heteroskedasticity (see the first scheme in Table 1). Note that the ordering of the equations in this scheme is to some extent arbitrary as no economic restrictions are imposed. In our empirical analysis, we use a model with two volatility regimes and order the equations such that the relative variances of the error terms of the *unrestricted* model correspond to the relative variances obtained for the conventional identification schemes. In particular, the equation with the largest relative variance will be placed as the fourth equation of the model and will be considered to be the interest rate equation because, for the conventional identification schemes discussed subsequently, the interest rate equation is the fourth equation and it has consistently the largest relative variance by far.

---

[3]We thank Belongia & Ireland for sharing their data set with us.

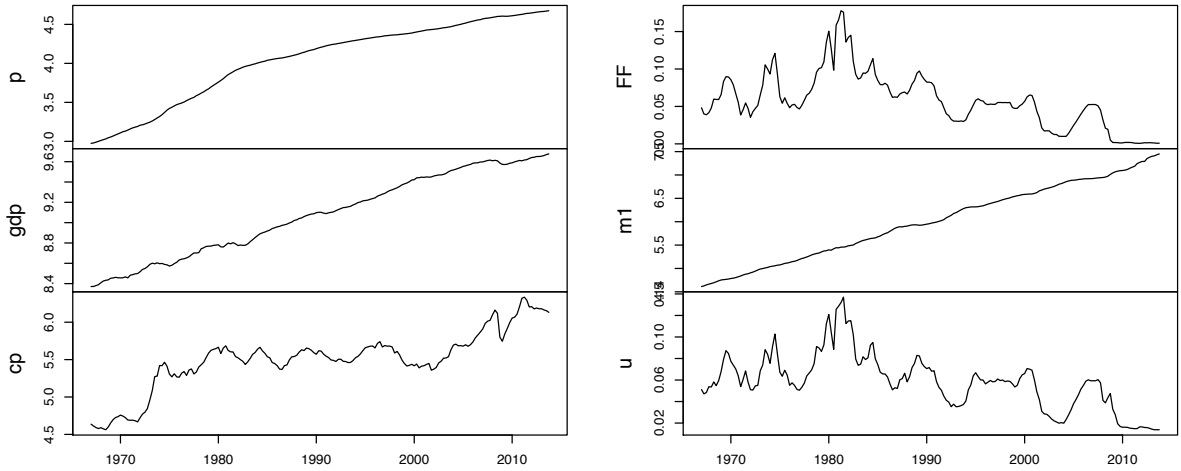

Figure 1: Time series data taken from Belongia & Ireland (2015).

A standard conventional identification scheme is a *recursive* model motivated by the work of Bernanke & Blinder (1992), Sims (1986, 1992) and others (see the second scheme in Table 1). It just-identifies the system in the conventional homoskedastic case. In the heteroskedastic case instead, the zero restrictions above the main diagonal are over-identifying and can be tested. This model identifies the monetary policy shock by imposing restrictions on the fourth row of the matrix $A_0$ such that the interest rate reacts to contemporaneous changes in the price level, output, and commodity prices.

In a monetary model the interest rate equation is typically set up as a Taylor rule which assumes that the interest rate reacts to inflation and the output gap. In our model comparison we include a benchmark model inspired by Leeper & Roush (2003) which is also discussed by Belongia & Ireland (2015). In Table 1 the identification of the $A_0$ matrix from this work is described as *Taylor Rule with Money*. In addition to standard Taylor rule variables such as output gap and inflation, the interest rate equation also contains the Divisia monetary aggregate, meaning that monetary policy reacts to changes in the money stock. Additionally, this model identifies the fifth and sixth shocks as money demand and money supply shocks, respectively. The restrictions imposed on the last two rows of $A_0$ over-identify the model and, hence, they can be tested in a conventional setting as well as in our heteroskedastic setting.

Belongia & Ireland are specifically interested in the role of the divisia money variable in the interest rate reaction function. Therefore they test two sets of over-identifying restrictions on the fourth row of $A_0$. First, they exclude the money aggregate from the Taylor rule by imposing the restriction $\alpha_{45} = 0$. This scheme is indicated as *Taylor Rule without Money* in Table 1. It corresponds to the standard monetary policy reaction function of Taylor (1993). Another specification considered by Belongia & Ireland (2015) is denoted as *Money-Interest Rate Rule* in Table 1. It assumes that the interest rate reacts contemporaneously only to the money aggregate. Such a rule was advocated by Leeper & Roush (2003) and used also in Leeper & Zha (2003) and

Table 1: Competing Monetary Policy Models

| Unrestricted | Recursive Scheme |
|---|---|
| $\begin{bmatrix} 1 & \alpha_{12} & \alpha_{13} & \alpha_{14} & \alpha_{15} & \alpha_{16} \\ \alpha_{21} & 1 & \alpha_{23} & \alpha_{24} & \alpha_{25} & \alpha_{26} \\ \alpha_{31} & \alpha_{32} & 1 & \alpha_{34} & \alpha_{35} & \alpha_{36} \\ \alpha_{41} & \alpha_{42} & \alpha_{43} & 1 & \alpha_{45} & \alpha_{46} \\ \alpha_{51} & \alpha_{52} & \alpha_{53} & \alpha_{54} & 1 & \alpha_{56} \\ \alpha_{61} & \alpha_{62} & \alpha_{63} & \alpha_{64} & \alpha_{65} & 1 \end{bmatrix}$ | $\begin{bmatrix} 1 & 0 & 0 & 0 & 0 & 0 \\ \alpha_{21} & 1 & 0 & 0 & 0 & 0 \\ \alpha_{31} & \alpha_{32} & 1 & 0 & 0 & 0 \\ \alpha_{41} & \alpha_{42} & \alpha_{43} & 1 & 0 & 0 \\ \alpha_{51} & \alpha_{52} & \alpha_{53} & \alpha_{54} & 1 & 0 \\ \alpha_{61} & \alpha_{62} & \alpha_{63} & \alpha_{64} & \alpha_{65} & 1 \end{bmatrix}$ |

| Taylor Rule with Money | Taylor Rule without Money | Money-Interest Rate Rule |
|---|---|---|
| $\begin{bmatrix} 1 & 0 & 0 & 0 & 0 & 0 \\ \alpha_{21} & 1 & 0 & 0 & 0 & 0 \\ \alpha_{31} & \alpha_{32} & 1 & \alpha_{34} & \alpha_{35} & \alpha_{36} \\ \alpha_{41} & \alpha_{42} & 0 & 1 & \alpha_{45} & 0 \\ -1 & \alpha_{52} & 0 & 0 & 1 & \alpha_{56} \\ -\alpha_{65} & 0 & 0 & \alpha_{64} & \alpha_{65} & 1 \end{bmatrix}$ | $\begin{bmatrix} 1 & 0 & 0 & 0 & 0 & 0 \\ \alpha_{21} & 1 & 0 & 0 & 0 & 0 \\ \alpha_{31} & \alpha_{32} & 1 & \alpha_{34} & \alpha_{35} & \alpha_{36} \\ \alpha_{41} & \alpha_{42} & 0 & 1 & 0 & 0 \\ -1 & \alpha_{52} & 0 & 0 & 1 & \alpha_{56} \\ -\alpha_{65} & 0 & 0 & \alpha_{64} & \alpha_{65} & 1 \end{bmatrix}$ | $\begin{bmatrix} 1 & 0 & 0 & 0 & 0 & 0 \\ \alpha_{21} & 1 & 0 & 0 & 0 & 0 \\ \alpha_{31} & \alpha_{32} & 1 & \alpha_{34} & \alpha_{35} & \alpha_{36} \\ 0 & 0 & 0 & 1 & \alpha_{45} & 0 \\ -1 & \alpha_{52} & 0 & 0 & 1 & \alpha_{56} \\ -\alpha_{65} & 0 & 0 & \alpha_{64} & \alpha_{65} & 1 \end{bmatrix}$ |

Note: The vector of variables at time $t$ is $y'_t = (p_t, gdp_t, cp_t, FF_t, m_t, uc_t)$. The fourth row of each matrix specifies the monetary policy reaction function and identifies the fourth shock as the monetary policy shock.

Sims & Zha (2006). The restrictions $\alpha_{41} = \alpha_{42} = 0$ are over-identifying in this model and they were not rejected by Belongia & Ireland (2015).

There are some major differences in the comparison of the models proposed in the present paper and the analysis conducted by Belongia & Ireland (2015). First of all, Belongia & Ireland did not allow for heteroskedasticity of the structural shocks. Consequently, they could only test the over-identifying specifications conditional on a set of just-identifying restrictions. Thus, their tests of the restrictions in the fourth row of $A_0$ are conditional on restrictions imposed in the other rows of $A_0$. By using the heteroskedasticity of the structural shocks we can test not only the restrictions imposed by Belongia & Ireland, but we can also test the restrictions in the fourth row and leave all other rows unrestricted, provided the fourth equation is identified through heteroskedasticity.

In order to investigate the importance of the money aggregate in the interest rate equation, we estimate the models mentioned above with heteroskedastic structural shocks. We estimate models with the full set of restrictions as presented in Table 1 and also test models in which all of the rows apart from the fourth row are left unrestricted. Finally, all of the models are confronted with a model solely identified by heteroskedasticity in which all the off-diagonal elements of matrix $A_0$ are estimated without any zero restrictions. Importantly, our approach allows us to statistically compare alternative monetary policy models that are not nested within one another. For instance, our Bayes factors allow us to compare the *recursive* model to each of the remaining monetary policy models despite the fact that neither of them is nested within the *recursive* one.

We fit VAR models of order $p = 4$, as in Belongia & Ireland (2015), to our full sample of quarterly data from 1967Q1 - 2013Q4 and also to a reduced sample from 1967Q1 - 2007Q4. Following Belongia & Ireland (2015), the shorter sample is considered because it excludes the financial crisis period which could affect the structure of monetary policy in the US and, hence, it might lead to distortions in our analysis.

Table 2: Estimated Relative Variances, $\hat{\omega}_2$, of Structural Shocks in Heteroskedastic Models

|  | Unrestricted | Recursive | Taylor Rule with Money | Taylor Rule without Money | Money-Interest Rate |
|---|---|---|---|---|---|
| **Sample period 1967Q1 - 2013Q4** | | | | | |
| $p_t$ | 4.268 | 4.448 | 4.747 | 4.193 | 4.315 |
|  | (1.081) | (1.216) | (1.301) | (1.160) | (1.212) |
| $gdp_t$ | 11.115 | 7.538 | 7.271 | 7.705 | 7.518 |
|  | (2.920) | (2.016) | (1.971) | (2.183) | (2.131) |
| $cp_t$ | 7.432 | 6.913 | 6.870 | 5.977 | 5.936 |
|  | (1.720) | (1.712) | (1.717) | (1.547) | (1.570) |
| $FF_t$ | 45.087 | 33.106 | 42.774 | 38.411 | 37.409 |
|  | (10.825) | (8.270) | (10.484) | (10.007) | (9.839) |
| $m_t$ | 7.800 | 8.995 | 6.373 | 7.040 | 7.171 |
|  | (1.970) | (2.517) | (1.673) | (1.909) | (1.973) |
| $uc_t$ | 3.806 | 5.678 | 5.934 | 5.760 | 5.842 |
|  | (0.991) | (1.592) | (1.616) | (1.586) | (1.591) |
| **Sample period 1967Q1 - 2007Q4** | | | | | |
| $p_t$ | 3.783 | 3.916 | 4.045 | 4.021 | 3.961 |
|  | (1.167) | (1.270) | (1.216) | (1.254) | (1.220) |
| $gdp_t$ | 9.859 | 6.884 | 6.766 | 7.126 | 7.197 |
|  | (3.273) | (2.034) | (1.991) | (2.151) | (2.229) |
| $cp_t$ | 3.694 | 3.507 | 3.485 | 3.307 | 3.321 |
|  | (0.992) | (0.959) | (0.940) | (0.918) | (0.921) |
| $FF_t$ | 45.560 | 31.044 | 38.188 | 31.209 | 33.711 |
|  | (13.908) | (8.497) | (10.773) | (8.892) | (9.598) |
| $m_t$ | 3.133 | 2.901 | 1.931 | 1.854 | 1.873 |
|  | (1.367) | (1.375) | (0.712) | (0.810) | (0.756) |
| $uc_t$ | 2.807 | 3.214 | 3.543 | 3.309 | 3.706 |
|  | (0.806) | (0.917) | (1.008) | (0.924) | (1.052) |

Note: The table reports posterior means and posterior standard deviations – in parentheses – of the structural shocks' relative variances, $\omega_2$, for the Markov-switching models with two states, $M = 2$.

*5.3. Assessing Heteroskedasticity*

We have fitted a two-state Markov process to capture possible changes in the volatility of the residuals. The estimated variances of the second regime relative to the variances in the first regime are shown in Table 2 for all the different identification schemes imposed on $A_0$. Thus the quantities in the table are the estimated elements of $\omega_2$. They are all distinct from one, indicating that the second regime indeed has different variances than regime 1.

The marginal posterior regime probabilities of the second volatility state are depicted in Figure 2. They show that roughly the first part and the last part of the sample constitute the second volatility regime and the middle part from the first half of the 1980s to the beginning of the financial crisis constitute the first volatility regime. There is also a short period around the change of the millennium which is assigned to the second volatility state in the longer sample. Since the relative variances of the second volatility state are greater than one, this regime is clearly a high-volatility state. With respect to the dating and the interpretation the first state resembles the *Greenspan* state found by Sims & Zha (2006) while the second one highly resembles the *Burns and Volcker* state (see, e.g., Woźniak & Droumaguet, 2015). A similar classification of the volatility states is also obtained when the model is fitted only to the reduced sample ending in 2007Q4. Thus, the assignment of states is reasonably similar in both samples and is, hence, not driven entirely by the potentially higher macroeconomic volatility during the financial crisis.

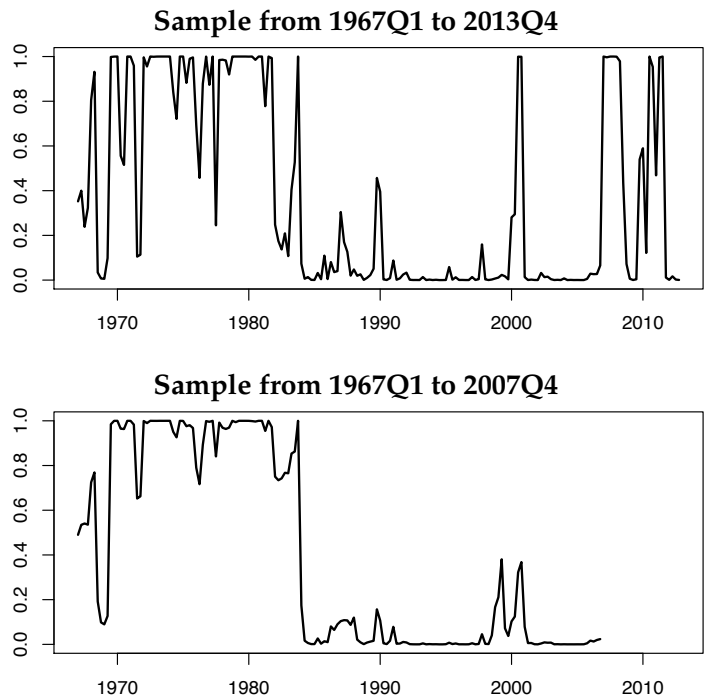

Figure 2: Marginal posterior probabilities of the second state for the best models with Markov-switching heteroskedasticity with 2 states.

In Table 2 the posterior standard deviations of the relative variances are also presented. Partly they are quite large. Therefore one may wonder whether the two estimated volatility states are really clearly distinct. This question can be answered by our formal statistical tools. In Table 3 we use SDDRs to assess whether the relative variances are actually 1.[4] Note that for our model with only two volatility regimes ($M = 2$), the $i^{\text{th}}$ structural shock is heteroskedastic if $\omega_{2,i} \neq 1$. It turns out that all results for the longer sample show strong support for the relative variances to be different from 1 which implies heteroskedasticity of all structural shocks. For the shorter sample the evidence is still strong that at least some of the relative variances are not equal to 1, while some others may not be different from 1. Apart from the unrestricted model, in all other models the evidence is strong that at most one shock is homoskedastic and, hence, we may still have full identification through heteroskedasticity, as discussed in Section 2. In any case, the evidence for two distinct volatility states is very strong for both samples because, to confirm distinct covariance matrices in the two states, it is enough that one of the relative variances differs from 1.

---

[4]In Appendix Appendix D we provide details on the precision of the estimated quantities. All results are sufficiently precise so as to ensure the qualitative validity of the results.

Table 3: Natural Logarithms of SDDRs for Assessing Heteroskedasticity

| | Unrestricted | Recursive | Taylor Rule with Money | Taylor Rule without Money | Money- Interest Rate |
|---|---|---|---|---|---|
| **Sample period 1967Q1 - 2013Q4** | | | | | |
| Hypothesis: $\omega_{2,i} = 1$ | | | | | |
| $p_t$ | -16.75 | -15.14 | -15.95 | -11.66 | -10.72 |
| $gdp_t$ | -80.55 | -36.14 | -35.98 | -26.93 | -29.59 |
| $cp_t$ | -52.62 | -27.78 | -31.45 | -27.2 | -23.63 |
| $FF_t$ | -541.29 | -393.51 | -542.65 | -406.74 | -373.75 |
| $m_t$ | -40.84 | -46.34 | -27.77 | -33.04 | -30.77 |
| $uc_t$ | -12.04 | -17.25 | -28.34 | -21.36 | -19.65 |
| Hypothesis: $\omega_{2,i} = 1$ for $i = 1, \ldots, N$ | | | | | |
| | -1058.07 | -849.44 | -981.9 | -773.1 | -838.41 |
| **Sample period 1967Q1 - 2007Q4** | | | | | |
| Hypothesis: $\omega_{2,i} = 1$ | | | | | |
| $p_t$ | -7.29 | -5.17 | -8.3 | -5.88 | -6.5 |
| $gdp_t$ | -44.09 | -27.07 | -31.50 | -28.49 | -31.07 |
| $cp_t$ | -9.86 | -7.29 | -7.89 | -6.55 | -6.67 |
| $FF_t$ | -406.27 | -280.94 | -369.25 | -287.87 | -341.29 |
| $m_t$ | -0.64 | -0.62 | 0.72 | 1.08 | 0.88 |
| $uc_t$ | -4.06 | -5.64 | -8.15 | -6.32 | -8.61 |
| Hypothesis: $\omega_{2,i} = 1$ for $i = 1, \ldots, N$ | | | | | |
| | -582.38 | -464.92 | -518.97 | -426.52 | -495.53 |

Note: The table reports natural logarithms of SDDRs for the hypothesis of homoskedasticity of individual structural shocks, as well as the hypothesis of joint homoskedasticity in the models with two volatility states, $M = 2$. Numbers in boldface denote SDDR values indicating very strong evidence against the hypothesis on a scale by Kass & Raftery (1995). The numerical standard errors for the SDDRs reported in this table are given in Appendix Appendix D.

*5.4. Assessing Identification*

These results clearly indicate that there is time-varying volatility in the data that can be used for identification purposes. It is therefore of interest to know whether there is sufficient heteroskedasticity to ensure a fully identified model. As discussed in the earlier sections, in a model with two states, full identification requires that all of the relative variances, $\omega_{2,n}$, $n = 1, \ldots, N$, are distinct. Again we can use SDDRs to investigate this identification condition. In Table 4 the relevant SDDRs are given. For both samples they provide strong support for at least some distinct relative variances. In particular, the SDDRs strongly indicate that $\omega_{2,4}$ is different from all other $\omega_{2,j}$, because all SDDRs related to $\omega_{2,4}/\omega_{2,j} = 1$ are very small. Thus, there is particularly strong support for the fourth equation to be identified. Recall that this equation is the interest rate equation which is of special interest in Belongia & Ireland (2015). Although there is also support for some of the other relative variances to be distinct, this support is less strong and there is a chance that only some of the equations of our model are identified by heteroskedasticity.

Table 4: Natural Logarithms of SDDRs for Assessing Identification

| | Hypothesis: $\omega_{2,i}/\omega_{2,j} = 1$ | | | | |
|---|---|---|---|---|---|
| **Sample period 1967Q1 - 2013Q4** | | | | | |
| i↓, j→ | 2 | 3 | 4 | 5 | 6 |
| 1 | -2.71 | -0.41 | **-22.85** | -0.45 | 1.02 |
| 2 | | 0.42 | **-7.44** | 0.49 | -3.42 |
| 3 | | | **-13.22** | 1.12 | -0.96 |
| 4 | | | | **-12.55** | **-24.35** |
| 5 | | | | | -0.97 |
| **Sample period 1967Q1 - 2007Q4** | | | | | |
| i↓, j→ | 2 | 3 | 4 | 5 | 6 |
| 1 | -1.92 | 1.01 | **-19.47** | 0.65 | 0.68 |
| 2 | | -2.47 | **-7.14** | -2.32 | -3.37 |
| 3 | | | **-22.77** | 0.63 | 0.69 |
| 4 | | | | **-18.87** | **-24.36** |
| 5 | | | | | 0.64 |

Note: The entries of the table are natural logarithms of SDDRs for the hypotheses of pairwise proportional changes in the volatility of the structural shocks in the models with two volatility states, $M = 2$, and unrestricted matrix $A_0$. Numbers in boldface denote SDDR values indicating very strong evidence against the hypothesis on a scale by Kass & Raftery (1995). The numerical standard errors for the SDDRs reported in this table are given in Appendix Appendix D.

*5.5. The Role of Money Revisited*

Given the results for the relative variances in our model, we can compare different identification schemes via their MDDs. For a range of models they are given in Table 5. Each row displays the MDDs for the five identification schemes listed in Table 1 for a different model setup. The model with the largest MDD in each row is highlighted in boldface.

Looking at the models identified through heteroskedasticity, models with divisia money in the interest rate equation have the largest MDDs for the longer sample according to the results in Table 5. The same is also true for the shorter sample when only zero restrictions are imposed on the interest rate equation. In contrast, the largest MDD is obtained for the scheme signified as *Taylor Rule without Money* when restrictions are imposed on all of the rows of matrix $A_0$. Note, however, that in this case the three schemes *Taylor Rule with Money*, *Taylor Rule without Money*, and *Money-Interest Rate* have almost identical MDDs such that the evidence in favor of a model without money in the interest rate equation is very weak at best.

The advantage of our setup is that we can deal also with models which are only partially identified in a conventional setting as they are compared in the second row of the two panels in Table 5. Thus, using heteroskedasticity we can compare models which impose restrictions on the equation which is of direct interest. We do not have to condition on the restrictions on the other rows of $A_0$, as in a conventional frequentist analysis. Clearly, if additional restrictions are imposed and then the restrictions are rejected in such a setup, it is unclear whether the restrictions of interest or the additional restrictions drive the rejection. In contrast, using heteroskedasticity it is possible

Table 5: Natural Logarithms of MDDs for Assessing Restrictions on $A_0$

|  | Unrestricted | Recursive | Taylor Rule with Money | Taylor Rule without Money | Money-Interest Rate |
|---|---|---|---|---|---|
| **Sample period 1967Q1 - 2013Q4** | | | | | |
| all restrictions | -1700.2 | -1657.6 | **-1632.5** | -1643.6 | -1644.6 |
| interest rate equation restricted | -1700.2* | -1695.9 | **-1687.8** | -1695.3 | -1688.1 |
| **Sample period 1967Q1 - 2007Q4** | | | | | |
| all restrictions | -1469.8 | -1428.6 | -1414.4 | **-1410.4** | -1411.2 |
| interest rate equation restricted | -1469.8* | -1457.6 | -1453.7 | -1453.6 | **-1442.1** |

Note: The table reports natural logarithms of marginal data densities for particular models. Numbers in boldface denote the largest values of the MDDs in rows. * denotes values that are copied from the row above. The numerical standard errors for the logarithms of the MDDs reported in this table are given in Appendix Appendix D.

to explicitly impose the restrictions only on the interest rate equation. The other parameters are identified by heteroskedasticity. Admittedly, this argument relies on full identification through heteroskedasticity which is not strongly supported for our data. However, identification of the interest rate equation is strongly supported confirming that the differences in MDDs are not only driven by our prior but reflect data properties. The last claim is based on the fact that we assume hierarchical prior distributions for the parameters for which the level of shrinkage is estimated. Thereby we leave considerable room for the data to speak. Overall our analysis supports the importance of divisia money in the interest rate equation.

## 6. Conclusions

This study considers structural VAR models with heteroskedasticity where the changes in volatility are driven by a Markov process. A full Bayesian analysis framework is presented for such SVAR-MSH models. A set of parametric restrictions for unique identification of the structural parameters through heteroskedasticity in these models is derived and Bayesian methods are presented for investigating the restrictions for global identification based on a Savage-Dickey density ratio. Moreover, a fast Markov Chain Monte Carlo sampler is developed for the posterior distribution of the structural parameters and a method for computing the marginal data density is provided which facilitates a full Bayesian model selection and model comparison.

SVAR models from a frequentist study by Belongia & Ireland (2015) are used to illustrate the Bayesian methods. Belongia & Ireland are interested in the role of a Divisia money aggregate in an interest rate reaction function. In the empirical illustration we compare our Bayesian methods to frequentist methods. It is shown that using heteroskedasticity for identification is beneficial and that this can be done in a Bayesian framework. In fact, our methods go beyond what is currently possible in a frequentist framework. In particular, in our Bayesian framework we can formally investigate conditions for identification of specific equations and shocks for which formal statistical tests are currently not available in a frequentist framework.

## Appendix A. Proof of Theorem 1

We first show the following matrix result.

**Lemma 1.** Given a sequence of positive definite $N \times N$ matrices $\Omega_m$, $m = 1, \ldots, M \geq 2$, let $C$ be a nonsingular $N \times N$ matrix and $\Delta_m = \text{diag}(\delta_{m,1}, \ldots, \delta_{m,N})$ be a sequence of $N \times N$ diagonal matrices such that
$$\Omega_m = C\Delta_m C', \quad m = 1, \ldots, M, \tag{A.1}$$
where $\Delta_1 = I_N$, the $N \times N$ identity matrix. Let $\delta_k = (\delta_{1,k}, \ldots, \delta_{M,k})$ be an $M$-dimensional vector. Then the $k^{\text{th}}$ column of $C$ is unique up to sign if $\delta_k \neq \delta_i \; \forall i \in \{1, \ldots, N\} \setminus \{k\}$.

**Proof:** The proof uses ideas from Lanne et al. (2010). Let $C_*$ be a matrix that satisfies
$$\Omega_m = C_* \Delta_m C_*', \quad m = 1, \ldots, M.$$
It will be shown that, under the conditions of Lemma 1, the $k^{\text{th}}$ column of $C_*$ must be the same as that of $C$, except perhaps for a reversal of signs. Without loss of generality it is assumed in the following that $k = 1$, because this simplifies the notation. In other words, it is shown that the first columns of $C$ and $C_*$ are the same possibly except for a reversal of signs.

There exists a nonsingular $N \times N$ matrix $Q$ such that $C_* = CQ$. Using condition (A.1) for $m = 1$, $Q$ has to satisfy the relation
$$CC' = CQQ'C'.$$
Multiplying this relation from the left by $C^{-1}$ and from the right by $C^{-1\prime}$ implies that $QQ' = I_N$ and, hence, $Q$ is an orthogonal matrix.

The relations
$$C\Delta_m C' = CQ\Delta_m Q'C', \quad m \in \{1, \ldots, M\},$$
imply
$$\Delta_m = Q\Delta_m Q' \quad \text{and, hence,} \quad Q\Delta_m = \Delta_m Q \quad \text{for } m \in \{1, \ldots, M\}.$$
Denoting the $ij^{\text{th}}$ element of $Q$ by $q_{ij}$, the latter equation implies that
$$q_{k1}\delta_1 = q_{k1}\delta_k, \quad k = 1, \ldots, N.$$
Hence, since $\delta_k$ is different from $\delta_1$ for $k = 2, \ldots, N$, we must have $q_{k1} = 0$ for $k = 2, \ldots, N$. Since, $Q$ is orthogonal, the first column must then be $(1, 0, \ldots, 0)'$ or $(-1, 0, \ldots, 0)'$, which proves Lemma 1. Q.E.D.

Using Lemma 1 the proof of Theorem 1 is straightforward.

*Proof of Theorem 1.* Consider the setup of Lemma 1 with $C = A_0^{-1}\Lambda_1^{1/2}$. Then the arguments in the proof of Lemma 1 show that any other admissible matrix $C$ is of the form $C_* = CQ = A_0^{-1}\Lambda_1^{1/2}Q$, where $Q$ is as in the proof of Lemma 1. Hence, $C_*^{-1} = Q'\Lambda_1^{-1/2}A_0$, which shows that $\Lambda_1^{-1/2}A_0$ and $Q'\Lambda_1^{-1/2}A_0$ have the same $k^{\text{th}}$ row. Multiplying the $k^{\text{th}}$ row of $\Lambda_1^{-1/2}A_0$ by $\sqrt{\lambda_{1,k}}$ gives the desired result. Q.E.D.

## Appendix B. Computational Details

*Appendix B.1. Notation and Likelihood Function*

Let the $N \times T$ matrix $Y = [y_1, \ldots, y_T]$ collect all the observations of the time series considered. Let $K = 1 + pN$ and define the $K \times 1$ vector $x_t$ as

$$x_t = \left(1, y'_{t-1}, y'_{t-1}, \ldots, y'_{t-p}\right)'.$$

It collects all the variables on the right-hand side (RHS) of equation (1). Moreover, let $X = [x_1, \ldots, x_T]$ be a $K \times T$ matrix, where the initial conditions $y_0, \ldots, y_{p-1}$ are treated as given and set to the first $p$ observations of the available dataset. Similarly, collect the structural errors in the matrix $U = [u_1, \ldots, u_T]$, and denote its $n^{\text{th}}$ row by $U_n$. Let $Y_m$, $X_m$, and $U_{n.m}$ denote matrices corresponding to the matrices $Y$, $X$, $U$, and $U_n$ collecting only the state specific columns for which $s_t = m$, $m \in \{1, \ldots, M\}$. The column dimension of these matrices is denoted by $T_m$ and $\sum_{m=1}^{M} T_m = T$. The $1 \times T$ vector $\mathbf{S} = (s_1, \ldots, s_T)$ is the realization of the hidden Markov process for periods from 1 to $T$. Define a $N \times K$ matrix $A = [\mu, A_1, \ldots, A_p]$ collecting the slope parameters and constant terms on the RHS of equation (1) and denote its $n^{\text{th}}$ row by $A_n$ which is a $1 \times K$ vector. For convenience we also denote by $\theta$ the vector of all the parameters of the model.

Using the previously defined notation, equation (1) can be written in matrix notation as

$$A_0 Y = AX + U, \tag{B.1}$$

and the $n^{\text{th}}$ row of (B.1) can be written as

$$A_{0.n} Y = A_n X + U_n, \tag{B.2}$$

for $n = 1, \ldots, N$.

Given the assumptions above and the conditional normality assumption in equation (2) for the structural errors of the SVAR-MSH model, the likelihood function is given by:

$$p(Y|\mathbf{S}, \theta) = (2\pi)^{-\frac{TN}{2}} |\det(A_0)|^T \left(\prod_{n=1}^{N} \lambda_{1,n}^{\frac{-T}{2}}\right) \left(\prod_{m=2}^{M} \prod_{n=1}^{N} \omega_{m,n}^{\frac{-T_m}{2}}\right) \times$$
$$\times \exp\left\{-\frac{1}{2} \sum_{m=1}^{M} \sum_{n=1}^{N} \lambda_{1,n}^{-1} \omega_{m,n}^{-1} [A_{0.n} Y_m - A_n X_m][A_{0.n} Y_m - A_n X_m]'\right\}, \tag{B.3}$$

where $\omega_{1,n} = 1$ for $n \in \{1, \ldots, N\}$. The likelihood function written in this form emphasizes the feature of the SVAR models that equations of the model can be analyzed one by one leading to a convenient form of the full conditional posterior distributions used in the Gibbs sampler.

*Appendix B.2. Gibbs Sampler*

*Sampling the variances of the structural shocks.* For given $Y$, $\mathbf{S}$, $A_n$ and $A_{0.n}$, each $\lambda_{1,n}$ is drawn independently, for $n \in \{1, \ldots, N\}$, from an $\mathcal{IG}2$ distribution:

$$\lambda_{1,n}|Y, \mathbf{S}, A_n, A_{0.n}, \omega_{m,n} \sim \mathcal{IG}2\left(\underline{a}_\lambda + 2T_1, \; \underline{b}_\lambda + \sum_{m=1}^{M} \omega_{m,n}^{-1} (A_{0.n} Y_1 - A_n X_1)(A_{0.n} Y_1 - A_n X_1)'\right).$$

Similarly, the relative variances $\omega_{m,n}$ are drawn independently, for $m \in \{2, \ldots, M\}$ and $n \in \{1, \ldots, N\}$, from the following $\mathcal{IG}2$ distribution:

$$\omega_{m,n}|Y, \mathbf{S}, A_n, A_{0.n}, \lambda_{1,n} \sim \mathcal{IG}2\left(\underline{a}_\omega + T_m, \underline{b}_\omega + \lambda_{1,n}^{-1}\left(A_{0.n}Y_m - A_n X_m\right)\left(A_{0.n}Y_m - A_n X_m\right)'\right).$$

*Sampling the structural matrix $A_0$.* To sample the posterior of the unrestricted elements of $A_0$ collected in the vector $\alpha$ (see equation (6)), rewrite the SVAR model from equation (1) as $\tilde{y}_t = \tilde{x}_t \alpha + u_t$, where $\tilde{y}_t = \left(y_t' \otimes I_N\right)q - Ax_t$, and $\tilde{x}_t = -\left(y_t' \otimes I_N\right)Q$. Then the likelihood function takes the following form:

$$p(Y|\mathbf{S}, \theta) = (2\pi)^{-\frac{TN}{2}} \prod_{t=1}^{T} \prod_{n=1}^{N} \lambda_{s_t,n}^{-\frac{1}{2}} |\det(A_0)|^T \exp\left\{-\frac{1}{2}\sum_{t=1}^{T}\left[\tilde{y}_t - \tilde{x}_t \alpha\right]' \operatorname{diag}(\lambda_{s_t})^{-1}\left[\tilde{y}_t - \tilde{x}_t \alpha\right]\right\}. \quad \text{(B.4)}$$

This likelihood function resembles a multivariate normal density function for $\alpha$, apart from the term $|\det(A_0)|^T$. This observation motivates the choice of the candidate-generating density in the following Metropolis- Hastings algorithm. Draw a candidate value, denoted by $\bar{\alpha}$, at the $s^{\text{th}}$ iteration from a multivariate $t$ distribution centered at the previous state of the Markov Chain, $\alpha^{(s-1)}$, with the scale matrix set to $P^*$ and the degrees of freedom parameter $v$, where $P^* = \left(\sum_{t=1}^{T} \tilde{x}_t' \operatorname{diag}(\lambda_{s_t})^{-1} \tilde{x}_t\right)^{-1}$. If $\alpha$ followed a multivariate normal distribution resembling the likelihood function from equation (B.4), then $P^*$ would be its covariance matrix. Then, compute $\delta = p(\bar{\alpha}|Y)/p(\alpha^{(s-1)}|Y)$, where $p(x|Y)$ is equal to the product of the likelihood function and the prior distribution evaluated at $x$, i.e., $p(Y|\mathbf{S}, x) p(x)$. Finally, draw $u$ from a uniform distribution on the interval $(0, 1)$ and set $\alpha^{(s)} = \bar{\alpha}$ if $u < \delta$ and $\alpha^{(s)} = \alpha^{(s-1)}$ otherwise. This Metropolis-Hastings algorithm is adjusted to the structural VAR identified through heteroskedasticity and in that respect it generalizes the algorithm by Canova & Pérez Forero (2015) maintaining its overall functionality.

*Sampling the autoregressive parameters.* The convenient form of the prior distribution and the likelihood function allow for sampling the constant term and the autoregressive parameters independently equation by equation from a multivariate normal distribution:

$$A_n'|Y, \mathbf{S}, A_{0.n}, \lambda_{1,n}, \omega_{m,n} \sim \mathcal{N}_K\left(A_{0.n}\overline{P}_n, \overline{H}_n\right),$$

for $n = 1, \ldots, N$, where

$$\overline{H}_n = \left[\lambda_{1,n}^{-1}X_1 X_1' + \lambda_{1,n}^{-1}\left(\sum_{m=2}^{M} X_m X_m'/\omega_{m,n}\right) + \widetilde{\underline{H}}^{-1}\right]^{-1}$$

and

$$\overline{P}_n = \left[\lambda_{1,n}^{-1}Y_1 X_1' + \lambda_{1,n}^{-1}\left(\sum_{m=2}^{M} Y_m X_m'/\omega_{m,n}\right) + \widetilde{\underline{P}}\widetilde{\underline{H}}^{-1}\right]\overline{H}_n.$$

Here $\widetilde{\underline{H}}$ is a diagonal matrix with the first element on the diagonal equal to $\gamma_\mu$ and the remaining ones equal to the diagonal of $\gamma_\beta \underline{H}$, and $\widetilde{\underline{P}} = [\mathbf{0}_{N\times 1}\ \underline{P}]$.

*Sampling the shrinkage parameters.* The shrinkage parameters $\gamma_\alpha$, $\gamma_\mu$ and $\gamma_\beta$ are sampled independently from the following $\mathcal{IG}2$ distributions:

$$\gamma_\alpha|Y,\alpha \sim \mathcal{IG}2\left(\underline{a} + r, \underline{b} + \alpha'\alpha\right),$$

$$\gamma_\mu|Y,\mu \sim \mathcal{IG}2\left(\underline{a} + N, \underline{b} + \mu'\mu\right),$$

$$\gamma_\beta|Y,A_0,\beta_n \sim \mathcal{IG}2\left(\underline{a} + pN^2, \underline{b} + \sum_{n=1}^{N}\left[\beta_n - A_{0.n}\underline{P}\right]'\underline{H}^{-1}\left[\beta_n - A_{0.n}\underline{P}\right]\right).$$

*Simulating the hidden Markov process.* In order to estimate the states of the hidden Markov process we apply the algorithms presented in Section 11.2 of Frühwirth-Schnatter (2006) that are based on the smoothing procedure by Chib (1996). We estimate a stationary hidden Markov process for the Markov-switching mechanism, and thus, we set the distribution $p(s_0|\mathbf{P})$ to the ergodic probabilities (see Frühwirth-Schnatter, 2006, Section 11.2).

*Sampling the transition probabilities matrix.* The transition probabilities $\mathbf{P}_m$, are sampled independently from an $M$-dimensional Dirichlet distribution given $\mathbf{S}$:

$$\mathbf{P}_m|\mathbf{S} \sim \mathcal{D}_M\left(\underline{e}_{m1} + N_{m1}(\mathbf{S}),\ldots,\underline{e}_{mM} + N_{mM}(\mathbf{S})\right),$$

for $m \in \{1,\ldots,M\}$. The parameters of the prior Dirichlet distributions are updated by the count of the transitions from the $i^{th}$ to the $j^{th}$ state given $\mathbf{S}$, denoted by $N_{ij}(\mathbf{S})$.

Estimation of the stationary Markov chain for the Markov-switching model requires a Metropolis-Hastings step because $p(s_0|\mathbf{P})$ is set to a vector of ergodic probabilities which depends on $\mathbf{P}$. For more details the reader is referred to Section 11.5.5 of Frühwirth-Schnatter (2006) or, for the case of a restricted matrix $\mathbf{P}$, Droumaguet, Warne & Woźniak (2017). Woźniak & Droumaguet (2015) use a restricted matrix of transition probabilities to model different pattens of heteroskedasticity of the structural shocks.

*Appendix B.3. Estimation of Marginal Data Densities*

To compute the posterior probabilities of alternative SVAR-MSH models we estimate the MDDs for a particular model, $\mathcal{M}$, defined as:

$$p(Y|\mathcal{M}) = \int_\Theta p(Y|\theta,\mathcal{M})p(\theta|\mathcal{M})d\theta,$$

where $\Theta$ denotes the parameter space of the parameter vector $\theta$, while $p(Y|\theta,\mathcal{M})$ and $p(\theta|\mathcal{M})$ denote respectively the likelihood function and the prior density for model $\mathcal{M}$ (below the conditioning on $\mathcal{M}$ is suspended and only used in the context of model comparison).

We apply a simple corrected arithmetic mean estimator proposed by Pajor (2016) that is based on the identity:

$$p(Y) = \frac{\mathbb{E}_\theta\left[p(Y|\theta)\mathcal{I}_O(\theta)\right]}{\Pr[O|Y]} = \frac{\Pr[O]}{\Pr[O|Y]}\mathbb{E}_\theta\left[p(Y|\theta)|O\right], \tag{B.5}$$

that is indexed by the subset $O \subseteq \Theta$, where $\mathbb{E}_\theta[.]$ denotes the expected value with respect to the prior distribution of $\theta$, $\mathbb{E}_\theta[.|O]$ denotes the conditional expected value given $O$, $\Pr[O]$ and $\Pr[O|Y]$ denote the prior and posterior probabilities, respectively, of set $O$ and $\mathcal{I}_O(\theta)$ denotes an indicator function that takes the value of one if $\theta \in O$, and zero otherwise.

Pajor (2016) shows that a consistent and unbiased estimator of the MDD in equation (B.5) is given by

$$\hat{p}(Y) = \frac{1}{J} \sum_{j=1}^{J} \frac{p\left(Y|\theta^{(j)}\right) p\left(\theta^{(j)}\right) \mathcal{I}_O\left(\theta^{(j)}\right)}{s\left(\theta^{(j)}\right)}, \tag{B.6}$$

where $\left\{\theta^{(k)}\right\}_{j=1}^{J}$ denotes a sample drawn from the importance density $s(.)$. In the estimator above $\hat{\Pr}[O|Y] = 1$ by defining the set $O$ as $\{\theta^* : p(Y|\theta^*) \geq c_O\}$, where $c_O$ is the minimum value of the likelihood function evaluated at the draws from the posterior distribution, as recommended by Pajor (2016). Moreover, following Pajor (2016) the importance density is set to a multivariate truncated normal density with the mean and covariance set to the posterior mean and posterior covariance of the parameters, respectively. The truncation is only active to ensure that $\theta^{(k)} \in \Theta$.

**Appendix C. Definition and Moments of the Inverse Gamma 1 Ratio Distribution**

This section specifies the inverse gamma 1 ratio distribution for a random variable that is defined as a ratio of two independent inverse gamma 1–distributed random variables. The probability density function as well as the moments of the distribution are established. These results may facilitate the computations if one prefers to parametrize the model in terms of the conditional standard deviations instead of conditional variances $\lambda$ and $\omega$ that were used in Section 2.

**Definition 2 (Inverse Gamma 1 Ratio distribution)** Let $x$ and $y$ be two strictly positive independent random variables distributed according to the following $\mathcal{IG}1$ distributions: $x \sim \mathcal{IG}1(a_1, b_1)$ and $y \sim \mathcal{IG}1(a_2, b_2)$, where $a_1$, $a_2$, $b_1$, and $b_2$ are positive real numbers and the probability density function of the inverse gamma 1 distribution is given by

$$f_{\mathcal{IG}1}(x; a, b) = 2\Gamma\left(\frac{a}{2}\right)^{-1} \left(\frac{b}{2}\right)^{\frac{a}{2}} x^{-(a+1)} \exp\left\{-\frac{1}{2}\frac{b}{x^2}\right\}. \tag{C.1}$$

Then, the random variable $z$, defined as $z = x/y$, follows the *Inverse Gamma 1 Ratio* ($\mathcal{IG}1\mathcal{R}$) distribution with the probability density function given by

$$f_{\mathcal{IG}1\mathcal{R}}(z; a_1, a_2, b_1, b_2) = 2B\left(\frac{a_1}{2}, \frac{a_2}{2}\right)^{-1} b_1^{\frac{a_1}{2}} b_2^{\frac{a_2}{2}} z^{a_2-1} \left(b_1 + b_2 z^2\right)^{-\frac{a_1+a_2}{2}}, \tag{C.2}$$

where $B(\cdot, \cdot)$ denotes the beta function. □

**Moments of the $\mathcal{IG}1\mathcal{R}$ distribution.** The expected value and the variance of the $\mathcal{IG}1\mathcal{R}$–distributed random variable $z$ are respectively given by

$$\mathbb{E}[z] = \left(\frac{b_1}{b_2}\right)^{\frac{1}{2}} \frac{B\left(\frac{a_1-1}{2}, \frac{a_2+1}{2}\right)}{B\left(\frac{a_1}{2}, \frac{a_2}{2}\right)} \quad \text{for} \quad a_1 > 2, \tag{C.3}$$

$$Var[z] = \frac{b_1}{b_2} \frac{a_2}{a_1 - 2} - \frac{b_1}{b_2} \left[\frac{B\left(\frac{a_1-1}{2}, \frac{a_2+1}{2}\right)}{B\left(\frac{a_1}{2}, \frac{a_2}{2}\right)}\right]^2 \quad \text{for} \quad a_1 > 4. \tag{C.4}$$

In general, the $k^{\text{th}}$ order non-central moment of $z$ is given by

$$\mathbb{E}\left[z^k\right] = \left(\frac{b_1}{b_2}\right)^{\frac{k}{2}} \frac{B\left(\frac{a_1-k}{2}, \frac{a_2+k}{2}\right)}{B\left(\frac{a_1}{2}, \frac{a_2}{2}\right)} \quad \text{for} \quad a_1 > 2k. \tag{C.5}$$

**Appendix D. Numerical Standard Errors for MDDs and SDDRs**

In Tables D.6 and D.7 we report the Numerical Standard Errors (NSEs) for the logarithms of the SDDRs for the assessment of the homoskedasticity and identification conditions reported in Tables 3 and 4, respectively. All of the values of the NSEs are small and show that our assessment measures are numerically stable. The values of the NSEs increase monotonically with increasing absolute value of the logarithm of the corresponding SDDRs. Nevertheless, the relative values of the NSEs to the logarithms of the SDDRs are negligible and do not affect the conclusions.

In Table D.8 we report the NSEs for the logarithms of the MDDs for the models that are reported in Table 5. These NSEs are greater in value than the NSEs for the logarithms of the SDDRs discussed above. However, the NSEs are smaller than the NSEs of the MDD estimator proposed for the heteroskedastic SVARs by Woźniak & Droumaguet (2015) that were computed for a similar simulation settings, but for a larger model ($N = 8$). These results allow us to state that our conclusions are reliable even in the case of the smallest difference between the MDDs for two models. The logarithm of the MDD for the *Taylor Rule without Money* model is significantly different from the corresponding value for the *Money-Interest Rate* model when all restrictions are imposed and for the sample ending in 2007. Still, the implied posterior probability of the former model is just over two times as large as the posterior probability of the latter one.

Table D.6: NSEs for Savage-Dickey Density Ratios for Assessing Heteroskedasticity

| | Unrestricted | Recursive | Taylor Rule with Money | Taylor Rule without Money | Money-Interest Rate |
|---|---|---|---|---|---|
| **Sample from 1967Q1 to 2013Q4** | | | | | |
| Hypothesis: $\omega_{2,i} = 1$ | | | | | |
| $p_t$ | 0.105 | 0.123 | 0.130 | 0.119 | 0.120 |
| $gdp_t$ | 0.466 | 0.245 | 0.216 | 0.270 | 0.251 |
| $cp_t$ | 0.237 | 0.280 | 0.245 | 0.216 | 0.216 |
| $FF_t$ | 1.889 | 1.382 | 1.554 | 1.635 | 1.492 |
| $m_t$ | 0.315 | 0.487 | 0.214 | 0.301 | 0.305 |
| $uc_t$ | 0.122 | 0.277 | 0.301 | 0.308 | 0.305 |
| Hypothesis: $\omega_{2,i} = 1$ for $i = 1, \ldots, N$ | | | | | |
| | 2.193 | 1.732 | 1.839 | 2.024 | 1.854 |
| **Sample from 1967Q1 to 2007Q4** | | | | | |
| Hypothesis: $\omega_{2,i} = 1$ | | | | | |
| $p_t$ | 0.084 | 0.112 | 0.089 | 0.103 | 0.097 |
| $gdp_t$ | 0.401 | 0.205 | 0.189 | 0.208 | 0.197 |
| $cp_t$ | 0.094 | 0.083 | 0.077 | 0.070 | 0.070 |
| $FF_t$ | 2.219 | 1.162 | 1.376 | 1.121 | 1.187 |
| $m_t$ | 0.133 | 0.091 | 0.012 | 0.013 | 0.014 |
| $uc_t$ | 0.047 | 0.081 | 0.107 | 0.082 | 0.119 |
| Hypothesis: $\omega_{2,i} = 1$ for $i = 1, \ldots, N$ | | | | | |
| | 2.722 | 1.280 | 1.546 | 1.276 | 1.329 |

Note: The table reports the Numerical Standard Errors for the ln SDDRs reported in Table 3 computed by the batch means method using 2000 batch means.

Table D.7: NSEs of Savage-Dickey Density Ratios for Assessing Identification

| **Sample from 1967Q1 to 2013Q4** | | | | | |
|---|---|---|---|---|---|
| i↓, j→ | 2 | 3 | 4 | 5 | 6 |
| 1 | 0.017 | 0.005 | 0.059 | 0.008 | 0.002 |
| 2 | | 0.006 | 0.034 | 0.006 | 0.026 |
| 3 | | | 0.044 | 0.001 | 0.013 |
| 4 | | | | 0.047 | 0.073 |
| 5 | | | | | 0.015 |
| **Sample from 1967Q1 to 2007Q4** | | | | | |
| i↓, j→ | 2 | 3 | 4 | 5 | 6 |
| 1 | 0.016 | 0.001 | 0.070 | 0.009 | 0.004 |
| 2 | | 0.017 | 0.035 | 0.047 | 0.033 |
| 3 | | | 0.062 | 0.010 | 0.005 |
| 4 | | | | 0.125 | 0.090 |
| 5 | | | | | 0.008 |

Note: The table reports the Numerical Standard Errors for the ln SDDRs reported in Table 4 computed by the batch means method using 2000 batch means.

Table D.8: NSEs for Marginal Data Densities

| | Unrestricted | Recursive | Taylor Rule with Money | Taylor Rule without Money | Money-Interest Rate |
|---|---|---|---|---|---|
| **Sample from 1967Q1 to 2013Q4** | | | | | |
| all restrictions | 0.138 | 0.142 | 0.145 | 0.138 | 0.141 |
| interest rate equation restricted | | 0.135 | 0.137 | 0.136 | 0.145 |
| **Sample from 1967Q1 to 2007Q4** | | | | | |
| all restrictions | 0.148 | 0.138 | 0.135 | 0.143 | 0.140 |
| interest rate equation restricted | | 0.142 | 0.134 | 0.132 | 0.134 |

Note: This table reports the NSEs of the estimates of the logarithms of the MDDs reported in Table 5. The NSEs are computed with the batch means method described in Perrakis, Ntzoufras & Tsionas (2014) based on 1000 batched means.